\documentclass{emulateapj}
\usepackage{apjfonts}
\usepackage{amsmath}
\usepackage{graphicx}
\usepackage{dcolumn}

\newcounter{ichi}
\newcounter{ni}
\newcounter{san}
\newcounter{yon}
\makeatletter
\newsavebox{\@parc@ption}
\def\parcaption#1{%
\sbox{\@parc@ption}{\shortstack[l]{#1}}%
>\setbox\@tempboxa\hbox{\csname fnum@\@captype\endcsname}%
\@tempdima\columnwidth \advance\@tempdima-\wd\@tempboxa
\@tempdimb\@tempdima 
\ifdim\wd\@parc@ption>\@tempdimb \@tempdima\@tempdimb
\else\@tempdima\wd\@parc@ption\fi
\sbox{\@tempboxa}{\parbox[t]{\@tempdima}{#1}}%
\caption{\usebox{\@tempboxa}}}
\makeatother

\slugcomment{}

\shorttitle{
BL Lac Objects as UHECR Sources
}
\shortauthors{Murase et al.}

\begin{document}

\title{Blazars as Ultra-High-Energy Cosmic-Ray Sources: Implications for TeV Gamma-Ray Observations}

\author{Kohta Murase\altaffilmark{1}, 
Charles D. Dermer\altaffilmark{2},
Hajime Takami\altaffilmark{3}, 
and Giulia Migliori\altaffilmark{4}
}

\altaffiltext{1}{Department of Physics, Center for Cosmology and Astro-Particle Physics, The Ohio State University, Columbus, OH 43210, USA}
\altaffiltext{2}{Space Science Division, Naval Research Laboratory, Washington, DC 20375, USA}
\altaffiltext{3}{Max Planck Institute for Physics, F\"{o}hringer Ring 6, 80805 Munich, Germany}
\altaffiltext{4}{Harvard-Smithsonian Center for Astrophysics, 60 Garden St., Cambridge, MA 02138, USA}

\begin{abstract}
The spectra of BL Lac objects and Fanaroff-Riley I radio galaxies are commonly explained by the one-zone leptonic synchrotron self-Compton (SSC) model.  Spectral modeling of correlated multiwavelength data gives the comoving magnetic field strength, the bulk outflow Lorentz factor and the emission region size. 
Assuming the validity of the SSC model, the Hillas condition shows that only in rare cases can such sources accelerate protons to much above ${10}^{19}$~eV, so $\gtrsim {10}^{20}$~eV ultra-high-energy cosmic rays (UHECRs) are likely to be heavy ions if powered by this type of radio-loud active galactic nuclei (AGN).  Survival of nuclei is shown to be possible in TeV BL Lacs and misaligned counterparts with weak photohadronic emissions. 
Another signature of hadronic production is intergalactic UHECR-induced cascade emission, which is an alternative explanation of the TeV spectra of some extreme non-variable blazars such as 1ES 0229+200 or 1ES 1101-232.  We study this kind of cascade signal, taking into account effects of the structured extragalactic magnetic fields in which the sources should be embedded.  
We demonstrate the importance of cosmic-ray deflections on the $\gamma$-ray flux, and show that required absolute cosmic-ray luminosities are larger than the average UHECR luminosity inferred from UHECR observations and can even be comparable to the Eddington luminosity of supermassive black holes.  
Future TeV $\gamma$-ray observations using the Cherenkov Telescope Array and the High Altitude Water Cherenkov detector array can test for UHECR acceleration by observing $>25$~TeV photons from relatively low-redshift sources such as 1ES 0229+200, and $\gtrsim$~TeV photons from more distant radio-loud AGN.
\end{abstract}


\section{Introduction}

Active galactic nuclei (AGN) with extended radio jets powered by super-massive black holes are among the most luminous objects in the low-redshift universe.  Since 2004, when the present generation of imaging atomspheric Cherenkov telescopes began to operate, the number of extragalactic sources detected at $\gtrsim 0.1 $~TeV (very-high energy; VHE) energies has grown rapidly, and is nearly fifty.\footnote{See tevcat.uchicago.edu/ and www.mpp.mpg.de/$\sim$rwagner/sources/}  The \textit{Fermi} Gamma-ray Space Telescope, now in its fourth mission year, is providing a wealth of new discoveries on $\gamma$-ray galaxies.  In the high-confidence clean sample of active galactic nuclei associations in the First \textit{Fermi} LAT AGN catalog \citep[1LAC;][]{abd101LAC}, more than 600 $\gamma$-ray blazars, divided about equally into BL Lac objects and flat spectrum radio quasars (FSRQs), were reported.  New classes of GeV $\gamma$-ray galaxies, e.g., radio-loud narrow line Seyfert galaxies~\cite{abd09a} and star-forming galaxies powered by supernovae rather than black holes \citep{abd10sb}, following closely the VHE detections of the starburst galaxies NGC 253~\citep{ace09} and M82~\citep{acc09c}, are now firmly established.  Moreover, the spectral energy distributions (SEDs) of radio galaxies detected at GeV and VHE, being misaligned by large ($\gtrsim 10^\circ )$ angles to the jet axis and thought to be the parent population of blazars in geometrical unification scenarios \citep{up95}, are helping to reveal the blazar jet geometry.  About 10 such sources are now detected with \textit{Fermi}~\cite{abd10ma}.
  
Radio-loud AGN detected at TeV energies consist mainly of high-synchrotron-peaked BL Lac objects, including the ultra-variable TeV blazars Mrk 421~\citep[$z$=0.031;][]{fos08}, Mrk 501~\citep[$z$=0.033;][]{alb07b} and PKS 2155-304~\citep[$z$=0.116;][]{aha07a}, and the apparently  non-variable TeV blazars 1ES 0229+200~\citep{aha07b} and 1ES 1101-232~\citep{aha07c}.
Extragalactic VHE $\gamma$-ray galaxies include several Fanaroff-Riley (FR) class I radio galaxies~\citep[Cen A, M87, NGC 1275;][respectively]{aha09a,aha06,ale11c}.  Cen A \citep{abd10cena}, M87~\citep{abd09m87}, and NGC 1275~\citep{abd09ngc1275,2010ApJ...715..554K,BA11} have also been detected at GeV energies.  This list also includes the head-tail radio galaxy IC 310~\citep{ale10,nsv10}, intermediate-synchrotron-peaked objects like 3C 66A~\citep{acc09a,ali09,abd11a} and BL Lac~\citep{alb07a,abd11b}, and the GeV luminous, high-redshift FSRQs 3C 279~\citep[$z$=0.538;][]{ale11a}, PKS 1510-089~\citep[$z$=0.361;][]{wag10}, and 4C +21.35~\citep[PKS 1222+216, $z$=0.432;][]{ale11b}.

The $\gamma$-ray data from weak-lined BL Lac objects and FR-I radio galaxies are generally well fit with the standard nonthermal electronic synchrotron self-Compton (SSC) relativistic jet model~\citep[e.g.,][]{mk97,tav98,kat06}, but the use of archival data for highly variable blazars gave large parameter uncertainties in the past.  With the recent simultaneous multi-wavelength data sets for many sources, accurate parameter estimation can be made, either from simple scaling results in the Thomson regime, or from detailed spectral calculations taking into account the Klein-Nishina (KN) effect that is relevant for high, UV/X-ray synchrotron-peaked blazars. 

The paper is organized as follows.  In Section 2, we assemble the derived parameter values obtained in various analyses of typical BL Lac objects and FR-I radio galaxies.  From these numbers, we obtain maximum energies of cosmic rays, and show that protons can be accelerated to $\gtrsim 10$~EeV energies only in a few radio galaxies and flares of BL Lac objects.  Then we explore the associated hadronic signatures expected from TeV blazars in the case where jets of BL Lac objects and FR-I radio galaxies are accelerators of ultra-high-energy cosmic rays (UHECRs, with energies above the ankle of $\approx {10}^{18.5}$~eV), assuming the validity of the leptonic SSC parameters inferred from rapidly variable AGN.  
We discuss observable signals produced in the source and calculate those generated outside the source from both $\gamma$ rays and UHECRs escaping from the jet accelerator and passing through the $\sim$~Mpc-scale regions of cosmic structure, magnetized clusters and filaments, and the larger $\sim 100$ Mpc-scale voids of intergalactic space.  For some apparently non-variable TeV blazars, where the one-zone synchrotron/SSC model typically requires extreme parameters for fits, the cascade radiation can be a crucial component of the high-energy radiation spectrum, depending on the strength of the intergalactic magnetic field (IGMF) in voids, as has recently been proposed to explain the $\gamma$-ray spectra of some extreme blazars such as 1ES 0229+200~\cite{ek10,ess10,ess11}. We focus on such cascade emissions in the VHE range in Section 3.  Using numerical calculations, we also demonstrate the importance of structured extragalactic magnetic fields (EGMFs) in clusters and filaments for future $\gamma$-ray detectability by the Cherenkov Telescope Array (CTA) and the High Altitude Water Cherenkov (HAWC) detector array.  Implications of this study and a summary are given in Section 4.  Throughout this work, the cosmological parameters are taken as $H_0=71~{\rm km}~{\rm s}^{-1}~{\rm Mpc}^{-1}$, $\Omega_m=0.3$, and $\Omega_\Lambda=0.7$.

\section{VHE Blazars and UHECRs}

Of the wide variety of source classes that could potentially accelerate UHECRs, including, for example, GRBs~\cite{wax95,vie95,mur08a}, fast rotating magnetars~\cite{aro03}, structure formation shocks in galaxy clusters~\cite{nma95,krj96,ino07}, and quasar remnants~\cite{BG99,lev00}, radio-loud AGN with jets seem privileged in that the most pronounced excess in arrival directions of UHECRs is positionally centered in the vicinity of the FR-I radio galaxy Centaurus A~\cite{abr08,abr09}.~\footnote{See Takami \& Sato 2009 for discussions on issues of UHECR anisotropy for protons, and Lemoine \& Waxman 2009; Abreu et al. 2011 for heavy nuclei.  Note also the potential contribution from the background Centaurus supercluster pointed out by Ghisellini et al. (2008).} 
FR-I radio galaxies, including their aligned counterparts (BL Lac objects), radiate a volume- and time-averaged emissivity of $\approx 10^{45}$--$10^{46}$~erg~Mpc$^{-3}$~yr$^{-1}$ in nonthermal $\gamma$ rays~\cite{dr10}, and FR-I radio galaxies are found within the $\approx 100$ Mpc Greisen-Zatsepin-Kuzmin radius.  If comparable power goes into the acceleration of UHECRs, then BL Lac objects and FR-I radio galaxies have more than sufficient emissivity to power the $\gtrsim 10$ EeV UHECRs, which require $\sim 10^{44}$~erg~Mpc$^{-3}$~yr$^{-1}$~\citep{wb99,bgg06,mt09}. 

\subsection{Two-Component Spectra of Blazars and Radio Galaxies}

In the framework of one-zone leptonic synchrotron/SSC models, the double-humped SED of blazars and radio galaxies (misaligned counterparts), plotted as log($\nu F_{\nu}$) vs. log($\nu$), can be well described by a low-energy synchrotron component and a high-energy inverse-Compton (IC) curve.  Each component is characterized by the peak synchrotron flux $\nu F_{\nu}^s$ at peak synchrotron photon energy $\varepsilon_s$, and the peak IC flux $\nu F_{\nu}^C$ at peak IC energy $\varepsilon_{C}$, respectively.  The spectrum of radio-loud AGN is often highly variable, and rises and decays with variability time $t_{\rm var}$.  Here, we define the variability time as the shortest timescale in which the flux shows a significant factor-of-two change, and assume co-spatiality of the highly variable emissions in the different energy bands, as indicated by correlated variability~\cite{ack11}. 

Here we consider such a one-zone model.  The emissions are produced by electrons in a spherical blob moving relativistically in a jetted geometry: the high-energy emission is produced via Compton scattering off the local synchrotron photons that are generated by the electrons in the blob.  The synchrotron (IC) luminosities at peak energy are $L_{\gamma}^{s(C)} \approx 4 \pi d_L^2 (\varepsilon_{s (C)} F_\varepsilon^{s (C)})$ and the Compton dominance parameter is defined as $A_C \equiv L_{\gamma}^{C}/L_{\gamma}^s \approx (\varepsilon_C F_{\varepsilon}^{C})/(\varepsilon_s F_{\varepsilon}^s)$.  The electron distribution is typically assumed to consist of power-law segments.  As long as both the $\nu F_{\nu}^s$ and $\nu F_{\nu}^C$ energy fluxes originate from electrons with the same mean comoving-frame energy $\gamma_b' m_e c^2$, the magnetic energy density can be rewritten in terms of $A_C$ as ${{B'}^2}/{8 \pi} \sim {L_{\gamma}^s}/{(4 \pi {R'}^2 \delta^4 c A_C)}$.  Then the Doppler factor is given  in this relativistic spherical blob formalism by the expression~\citep{gmd96}
\begin{equation}
\delta \sim \frac{{3^{1/2} (L_{\gamma}^{s})}^{1/4} {(\varepsilon_C/m_e c^2)}^{1/2}}{2^{3/4} c^{3/4} t_{\rm var}^{1/2} A_C^{1/4} B_{\rm Q}^{1/2} (\varepsilon_s/m_e c^2)}\;,
\end{equation}
and the comoving magnetic field is 
\begin{eqnarray}
B^\prime &\sim& (1+z) \frac{2^{11/4} c^{3/4} t_{\rm var}^{1/2} A_C^{1/4} B_{\rm Q}^{1/2} {(\varepsilon_s/m_ec^2)}^3}{3^{3/2} {(L_{\gamma}^s)}^{1/4} {(\varepsilon_C/m_e c^2)}^{3/2}} \nonumber \\ 
&\sim& (1+z) \frac{4 B_{\rm Q} (\varepsilon_s/m_e c^2)}{3 \delta (\varepsilon_C/\varepsilon_s)} \;,
\end{eqnarray}
provided the Compton scattering takes place in the Thomson regime, which applies when $\delta \gtrsim \delta_T = 2 \sqrt{3}\sqrt{\varepsilon_s \varepsilon_c/m_e^2 c^4} (1+z)$.\footnote{The inequality is derived from $4 \gamma_b' \varepsilon_s (1+z)/\delta \lesssim m_e c^2$ and $\varepsilon_C \approx (4/3) {\gamma_b'}^2 \varepsilon_s$.}  Here the critical magnetic field is defined as $B_{\rm Q} \equiv m_e^2 c^3/e \hbar \simeq 4.4 \times {10}^{13}~\rm G$ \citep[e.g.,][]{bp87}.  The above equations are derived by using the common relations $\varepsilon_s/m_ec^2 \approx {\delta} (B^\prime/B_{\rm Q})  {\gamma_{b}'}^2 /(1+z)$, and $\varepsilon_c/m_ec^2 \approx (4/3)  {\gamma_{b}'}^2 \varepsilon_s$ (i.e., the typical fluid-frame Lorentz factor of electrons radiating near the peak synchrotron and SSC frequencies is $\gamma_{b}' \approx  (\sqrt{3}/2) \sqrt{\varepsilon_c/\varepsilon_s}$ in the Thomson limit) \citep[e.g.,][]{sik09}. 
 
For high-peaked BL Lac objects, the Compton scattering often occurs in the KN regime, where more detailed modeling is required.  Nevertheless, equations~(1) and (2) demonstrate that source parameters such as $\delta$ and $B'$ can be determined from the double-humped SED.\footnote{Other parameters such as the acceleration efficiency $\eta$ ($t_{\rm acc}' = \eta \gamma' m_e c^2/(e B' c)$) depend on details of the electron distribution.  For example, for the simple power-law injection of accelerated particles, large values of $\eta$ are often suggested~\cite{it96}, which may not be the case for injection with multiple power-law segments.} 
 
Table~1 gives measured and inferred properties for blazars and radio galaxies with good multiwavelength coverage that are well described by a synchrotron/SSC model, where derived values of magnetic fields and Doppler and Lorentz factors based on detailed synchrotron and SSC modeling are taken from the literature (rather than using equations~1 and 2).  In Table 2, we also show parameters derived with equations~(1) and (2) (only when $\delta \gtrsim \delta_T$ is satisfied).
For blazars, we assume that $\Gamma\approx \delta$, whereas values of the angle of the jetted emission with respect to the observer inferred from observations are considered for radio galaxies.  Also, $R^\prime \approx c t^\prime_{\rm var} = c \delta t_{\rm var}/(1+z)$ is used.

The SED modeling has much improved thanks to the constantly increasing multi-wavelength coverage, which also allows simultaneous multi-band observations.  Nevertheless, there is still some degree of scatter among parameter sets obtained by different groups, partially related to unavoidable parameter-degeneracy with respect to the observables.  
In some cases, there is even significant scatter between derived parameters for the same state of a specific source (Cen A) or between different states of the same source (Mrk 501).  In particular,  the whole SED of Cen A up to the TeV-band cannot be fitted with a unique parameter set.  Abdo et al. (2010d) show different models and parameter values for the same data, which reflect the limitations of fitting single epoch, single component SEDs for derivations of source parameters. However, our conclusions are not affected by uncertainties of the poorly constrained parameters, as we discuss in more detail below.

\begin{table*}[hbt]
\rotatebox{90}{\begin{minipage}{\textheight}
\begin{center}
\begin{scriptsize}
\caption{Measured and inferred properties of VHE blazars and radio galaxies}
\begin{tabular}{|c|l|c|c|c|c|c|c|c|c|c|c|c|c|c|} \hline
ID & \multicolumn{1}{|c|}{Source} & $z$ 
& Epoch 
&$t_{\rm var}$  
& $\delta^{(a)}$ 
&$\Gamma^{(a)}$/$\theta_{\rm obs}^{(a,b)}$  & $\gamma_{b}^{\prime (a)}$  
& $\varepsilon_s^{(a)}$ & $\nu_sF_{\nu}^{s(a)}[10^{-10}]$ & 
$R'^{(a)}[10^{15}]$ & $B'^{(a)}$ 
& $\varepsilon_C^{(a)}$ & $\nu_C F_{\nu}^{C(a)} [10^{-10}]$ & Ref. \\
& & & 
& [s] 
& &/[deg]  & 
& [$m_ec^2$] & [erg cm$^{-2}$ s$^{-1}$] & 
[cm] & [G] 
& [$m_ec^2$] & [erg cm$^{-2}$ s$^{-1}$] & \\ \hline
1 & CenA(core) & 0.00183 
& 2009 
& $\leq 1.0 \times 10^{5}$ 
& 1.0-3.9 & 2.0-7.0/15-30 & $(0.8-400)\times10^3$  
& $(0.8-4000) \times 10^{-7}$ & 0.09-4.5 & $3.0-11.0$ 
& 0.02-6.2 
& 0.17-($8.3\times10^5$) & 0.025-8.5 & 1 \\
2 & M87 & 0.00436 
& 2009
& $1.7 \times 10^{5}$  
& 3.9 & 2.3/10 
& $4\times10^3$ 
 & $1.6 \times 10^{-7}$ & 0.06 & $14.0$ 
& 0.055  
& 18.6 & 0.068 & 2 \\
3 & NGC1275 & 0.0179 
& Oct. 2010$^{d}$ 
& $8.6 \times 10^{4}$ 
& 2.3 
& 1.8/25 & 960  
& $2.4 \times 10^{-3}$ & 0.9 
& $2 \times 10^{3}$ & 0.05  
& $2.9 \times 10^{3}$ 
& $0.3$ &3 \\
 4 & NGC6251 &0.024 
 & -- 
 & -- 
 &2.4 &2.4/25 &$2\times10^4$
 &$6.5\times10^{-7}$ &0.012 &120 &0.037 
 &7.3 &0.047 &4\\
5 & Mrk421 & 0.03 
& 19 March 2001 
& $1.0 \times 10^{3}$ 
& 80 & 80 
& 9.3$\times10^4$ 
& $0.005$ & 7.4 & 3.0 & 0.048  
& $8.1\times10^{4}$ & 7.0 & 5 \\
6 & Mrk501(h.$^{(c)}$,1997) & 0.0337 
& 16 April 1997 
& $7 \times 10^{3}$ 
& 14-20 
& 14-20 & $(7-300) \times 10^{4}$  
& 0.3-0.5 & 8.0-8.5
& $1.0-5.0$ & 0.15-0.8  
& $(1.4-2.6) \times 10^{6}$ 
& 2.9-3.4 & 6,7,8 \\
7 & Mrk501 (l.$^{(c)}$,1997) & 0.0337 
& 7 April 1997 
& -- 
& 15 & 15 
& $6 \times 10^{5}$
& $0.002$ & 0.63 & $5.0$ & 0.8 
& $4.4 \times 10^{5}$ & 0.4 & 6\\
8 & Mrk501 (l.$^{(c)}$,2007)& 0.0337 
& 2007 
& -- 
& 25 & 25 & $1 \times 10^{5}$  
& 0.002 & 0.63 & $1.0$ & 0.31  
& $4.4 \times 10^{5}$ & 0.4 & 9 \\
9 & Mrk501 (l.$^{(c)}$,2009) & 0.0337 
& 2009 
&$3.5\times10^5$  
& 12-25 & 12-25 & $(6-90) \times 10^{4}$ 
 & 0.002 & 0.55-0.63 & $1.0-130$ & 0.015-0.34
& $(1.3-4.4) \times 10^{5}$ & 0.3-0.4 & 7,10,11 \\
10  & 1ES1959+650(h.$^{(c)}$) & 0.047 
&Sept2001-May2002
&$(2.2-7.2)\times10^4$ 
&18-20  &14-20 &$4-5\times10^4$
 &$(0.07-8)\times10^{-3}$  &1.0-3 &5.8-9  &0.04-0.9 
&$8\times10^{5-6}$ &0.2-2 &12,13\\
11  & 1ES1959+650(l.$^{(c)}$) & 0.047 
&23-25 May2006  
&$8.64\times10^4$  
&18
&18 &$5.7\times10^4$ 
& $0.003$ &2.6 & $7.3$ & 0.25-0.4  
& $1.2 \times 10^{5}$ 
& 0.22 & 14,15 \\
12 & PKS2200+420/BL Lac & 0.069 
& -- 
& -- 
& 15 & 15 & 900.0 
 & $5.3 \times 10^{-7}$ & 0.76 & $2.0$ & 1.4 
 & 1.6
 & 0.4 & 14\\
13 & PKS2005-489 & 0.071 
& -- 
& -- 
& 22 & 22 & $1.3 \times 10^{4}$ 
 & $4.7 \times 10^{-5}$ & 1.5 & $8.0$ & 0.7 
& $3.6\times10^3$ & 0.07 & 14 \\
14 & WComae & 0.102 
&7-8 June 2008
 & 5400 
 & 20-25 & 20-25 
& $(1.5-20) \times 10^{4}$  
& $8.0 \times 10^{-5}$ & 0.4 
& 3.0 & 0.24-0.3 
& $8.1\times10^3$
& 0.15 & 14,16 \\
15 & PKS2155-304 & 0.116 
& 28-30 July 2006 
& 300 
& 110 
& 110 
& $4.3 \times 10^{4}$ 
& $4 \times 10^{-4}$ & 2.13 & 0.86 
& 0.1 
 & $9.7 \times 10^{5}$  
& 20.0 & 5 \\
\hline
\end{tabular}
\end{scriptsize} \\
\end{center}
\vspace{0.3cm}
$^{(a)}$: parameter value from the SED modeling in literature (see references); $^{(b)}$:
for blazar sources $\delta\approx\Gamma$ and $\theta_j\approx1/\Gamma$; $^{(c)}$: high (h.) and low (l.) state;
References: 1- Abdo et al.(2010d) (see Figure 5 and Table 2 in the paper for the different models), 2- Abdo et al. (2009c), 3- Abdo et al. (2009b) (see also Brown \& Adams 2011), 4- Migliori et al. (2011), 5- Finke et al. (2008), 6- Pian et al. (1998), 7- Acciari et al. (2011), 8- Katarzynski et al. (2001), 9- Albert et al. (2007b), 10- Anderhub et al. (2009), 11- Abdo et al. (2011c), 12- Tagliaferri et al. (2003), 13- Krawczynski et al. (2004), 14- Tavecchio et al. (2010), 15- Tagliaferri et al. (2008), 16- Acciari et al. (2009b).
\end{minipage}
}
\end{table*}

\subsection{Implications for UHECR Acceleration}

The Hillas condition~\cite{hil84} limits the maximum accelerated energy of ions with charge $Z$ to
\begin{equation}
E_{A}^{\rm max} \approx Z e B' \Gamma R', 
\label{ea}
\end{equation}
in order that the particle Larmor radius is smaller than the characteristic size scale $R^\prime \lesssim c t^\prime_{\rm var} = c \delta t_{\rm var}/(1+z)$.  The inequality is replaced by an equality in our estimates.  
Equation~(\ref{ea}) can be rewritten using the Thomson-limit relations given above when $\delta \approx \Gamma$, in which case one gets 
\begin{equation}
E_{A}^{\rm max} \sim Z e \frac{4{(L_{\gamma}^{s})}^{1/4} t_{\rm var}^{1/2} B_{\rm Q}^{1/2} \varepsilon_s}{3 c^{3/4} A_C^{1/4} \varepsilon_{C}^{1/2}}. 
\label{ea2}
\end{equation}
Notice that $E_A^{\rm max}$ is defined in the cosmic rest frame at UHECR production. 
 
In Table~2, the maximum proton energy is estimated using equation~(3) and parameters given in Table~1, which in turn are based on the results of synchrotron/SSC model fits for these sources found in dedicated modeling papers (rather than using equations~1 and 2).  Here, note that other losses and details of the acceleration process could limit the maximum particle energy further.  For the cases considered in Table~1, it is barely possible to accelerate protons up to $\sim {10}^{20}$~eV, whereas Fe nuclei could easily each reach $\gtrsim {10}^{20}$~eV provided that they can survive photodisintegration.  Alternately, $\gtrsim {10}^{20}$~eV proton acceleration could occur transiently during rare bursts or flares~\citep{der09,mt09}, though according to Table~2 it might still be difficult even for bright flares from Mrk 501 and PKS 2155-304.
As noticed above, there is the large scatter in parameters due to uncertainties.  However, even with the allowed spread in the parameter values (see Table~2), this conclusion seems robust.

\begin{table*}[hbt]
\begin{center}
\begin{scriptsize}
\caption{Inferred properties of VHE blazars and radio galaxies} 
\begin{tabular}{|c|l|c|c|c|c|c|c|c|c|c||c|}
\hline
  \multicolumn{1}{|c|}{ID} &
  \multicolumn{1}{c|}{Source} &
  \multicolumn{1}{c|}{d$_L$} &
  \multicolumn{1}{c|}{$L^s_{\gamma}[10^{45}]$} &
  \multicolumn{1}{c|}{$L^C_{\gamma}[10^{45}]$} &
  \multicolumn{1}{c|}{$A_C$} &
  \multicolumn{1}{c|}{$\gamma_b^{\prime\,(e)}$} &
  \multicolumn{1}{c|}{$\delta_T^{(b)}$} &
  \multicolumn{1}{c|}{$\delta^{(c)}$} &
  \multicolumn{1}{c|}{$R^{\prime\,(d)}[10^{15}]$} &
  \multicolumn{1}{c||}{$B^{\prime\,(a)}$} &
  \multicolumn{1}{c|}{$E^{\rm max\,(f)}_{A}/Z[10^{19}]$} \\
  \multicolumn{1}{|c|}{} &
  \multicolumn{1}{c|}{} &
  \multicolumn{1}{c|}{[Mpc]} &
  \multicolumn{1}{c|}{[erg cm$^{-2}$ s$^{-1}$]} &
  \multicolumn{1}{c|}{[erg cm$^{-2}$ s$^{-1}$]} &
  \multicolumn{1}{c|}{} &
  \multicolumn{1}{c|}{} &
  \multicolumn{1}{c|}{} &
  \multicolumn{1}{c|}{} &
  \multicolumn{1}{c|}{[cm]} &
  \multicolumn{1}{c||}{[G]} &
  \multicolumn{1}{c|}{[eV]} \\
\hline
  1 & CenA(core) & 3.7 & $(0.15-7.3)\times10^{-4}$ &
  $(0.04-14)\times10^{-4}$ & 0.3-1.9 & 890-2.1$\times10^4$ &9.9-($6\times10^{-4})$ & 0.12-3.7
  & 3.0-12
  &  0.02-9.1 &0.01-4\\

  2 & M87 & 16.7 & $2.0\times10^{-4}$ & $2.3\times10^{-4}$ & 1.1
  & $9.3\times10^3$ &0.006 & 2.7
  & 20
  & 0.021 & 0.05\\
  
  3 & NGC1275 &75.3  &0.06  &0.02  &0.35 & $960$ &0.005 & -- 
  & --
  & --
  &5\\

  4 & NGC6251 &104 &$2\times10^{-3}$ &$6.6\times10^{-3}$ &3.3 & $2.9\times10^3$ &0.007
  & -- & --  & --  &0.3\\

  5 & Mrk421 & 130.0 & 1.5 & 1.4 & 0.95 & $3.4\times10^3$ &74 & -- 
  & -- 
  & -- & 0.3\\

  6 & Mrk501 (h.$^{(g)}$,1997) & 146.0 & 2.0-2.2 & 0.7-0.9 & 0.36-0.41 &  (1.4-2.5)$  \times10^3$  &(3.0-3.5)$\times10^3$ & -- 
  & -- 
  & -- & 0.1-2\\

  7 & Mrk501 (l.$^{(g)}$,1997) & 146.0 &0.2-0.4 &0.1-0.2 &0.44-0.63 & (0.08-1.3)$\times10^4$ &100-1700 & -- &--
  & -- &2\\

  8 & Mrk501 (l.$^{(g)}$,2007) & 146.0 & 0.2 & 0.1 & 0.63 &   $1.3\times10^4$ &100 &-- &-- & --
  & 0.2\\

  9 & Mrk501 (l.$^{(g)}$,2009) & 146.0 &0.1-0.2 &0.08-0.1  &0.55-0.63 & $(0.7-1.3)\times10^4$ & 58-100 & --
  &--
  & -- &0.2-0.7\\

  10 & 1ES1959+650(h.$^{(g)}$) & 206 & 0.5-1.5 & 0.1-1.1 & 0.2-0.8 & $(2.7-9.5)\times10^4$ &27-910 &-- &-- 
  & -- & 0.1-3\\

  11 & 1ES1959+650(l.$^{(g)}$) & 206 & 1.3 & 0.1 & 0.08 & 6600 
  &66  & --
  & --
  & -- & 1-2\\

  12 & PKS2200+420/BL Lac & 307.0 & 0.8 & 0.45 & 0.53 & 2.8$\times10^3$  &0.006 
  & -- &-- & -- & 1\\

  13 & PKS2005-489 & 316.0 & 1.8 & 0.07 & 0.04 & 7.6$\times10^3$ &1.5 
  & -- & -- & -- & 4
  \\

  14 & WComae & 464.0 & 1.0 & 0.38 & 0.38 & 8.7$\times10^3$ &3.1 & 7.2 
  & 3.0-3.7 &
  2.1-2.6 & 0.4-0.7\\

  15 & PKS2155-304 & 533.0 & 7.2 & 68 & 9.4 &  1.3$\times10^4$  &24 & --
  & --
  & -- 
  & 0.3\\
\hline
\end{tabular}
\end{scriptsize} \\
\end{center}
\vspace{0.3cm}
 $^{(a)}$: obtained from equation (2);
 $^{(b)}$: $\delta_T=2\sqrt{3}\sqrt{\varepsilon_C \varepsilon_s/m_e^2 c^4}(1+z)$;
  $^{(c)}$: obtained from equation (1); $^{(d)}$: calculated assuming
  $R^\prime \approx c t^\prime_{\rm var} = c \delta t_{\rm
  var}/(1+z)$; 
  $^{(e)}$:
 $\gamma_b^{\prime}\approx\sqrt{3}/2\sqrt{\varepsilon_C/\varepsilon_s}$; $^{(f)}$: obtained from equation (3) using $B^\prime$, $\Gamma$, and $R^\prime$ reported in Table 1;
 $^{(g)}$: high (h.) and low (l.) state.
\end{table*}

A similar conclusion is also reached when considering luminosity requirements for BL Lac objects and FR-I radio galaxies~\citep{dr10,ghi10}.  In Ghisellini et al. (2010), physical parameters were obtained via spectral modeling using their one-zone leptonic model of all blazars with known redshift detected by the \textit{Fermi} satellite during its first 3-month survey.  The inferred magnetic luminosity of BL Lac objects is typically $L_{B} \sim {10}^{46}\delta_1^2~{\rm erg}~{\rm s}^{-1}$ and almost all of them satisfy $L_{B} \lesssim 2 \times {10}^{47} \delta_1^2~{\rm erg}~{\rm s}^{-1}$, where $\delta_1 = \delta / 10$.  On the other hand, the required magnetic luminosity for UHECR acceleration to $10^{20} E_{A,20}^{\rm max}$ eV is $L_B \approx 2 \times {10}^{47}\Gamma_1^2 {(E_{A,20}^{\rm max})}^2 Z^{-2}~{\rm erg}~{\rm s}^{-1}$ where $\Gamma_1 = \Gamma / 10$.  Hence, it also suggests difficulties in acceleration of $\gtrsim {10}^{20.5}$~eV protons in typical BL Lac objects, though the simple SSC model cannot be so simply applied to lower-peaked BL Lac objects where values of  $B'$ and $\delta$ are not well-defined due possibly to external Compton scattering components. 

If BL Lac objects and FR-I radio galaxies, as has often been considered~\citep[e.g.,][]{tt01,bgg02,der09}, accelerate the UHECRs, then an ultra-high-energy (UHE) proton origin of the highest-energy cosmic rays is disfavored from spectral modeling if the standard synchrotron/SSC model is correct.  Heavier nuclei can, however, be accelerated up to ultra-high energies. The composition of UHECRs is an open question, with both proton and heavy-ion dominated compositions having been claimed to be compatible with HiRes~\cite{abb10} and the Pierre Auger Observatory (PAO)~\cite{abr10} data, respectively.  As seen here, the standard model for $\gamma$-ray emission from BL Lac objects and FR-I radio galaxies suggests a transition from proton to heavy-ion dominated composition at $\sim  ({10}^{18}-{10}^{19})$~eV. 

We have assumed that $\gamma$ rays from TeV blazars and radio galaxies are from  leptonic Compton scattering in order that synchrotron theory can be used to derive the various parameters.  Thus the hadronic $\gamma$-ray flux must be considerably smaller for a consistent interpretation. 
Sufficiently high-energy protons and nuclei interact with synchrotron photons in the jet via the photomeson process, with photopion production efficiency $f_{p \gamma}$ for cosmic-ray protons estimated to be~\citep[e.g.,][]{mb10}
\begin{equation}
f_{p \gamma} \simeq 
2.3 \times {10}^{-4}~\left(\frac{2.5}{1+\alpha} \right) L_{\gamma,46}^s t_{\rm var,4}^{-1} \delta_1^{-4} {\left( \frac{1~\rm keV}{\varepsilon_s} \right)} {\left( \frac{E_p}{E_p^b} \right)}^{\alpha-1},
\label{fpgamma}
\end{equation}
where $E_p^b \simeq 1.6 \times {10}^{16}~{\rm eV}~{(\varepsilon_s/1~\rm keV)}^{-1} \delta_1^2 {(1+z)}^{-2}$ is the typical energy of a proton that interacts with a photon with $\varepsilon_s$ (where the proton energy is here defined in the observer frame). Also, $\alpha$ is the photon index at energies below or above $\varepsilon_s$.  For $\alpha \approx 1.5$, which is typical of BL Lac objects at $E_p > E_p^b$, the photomeson production efficiency at $\sim {10}^{19}$~eV becomes of order $f_{p \gamma} \sim 6 \times {10}^{-3}$, which suggests that the photomeson process is inefficient for this kind of blazar (though it could be more efficient for low-peaked BL Lac objects and FSRQs).  The efficiency can also be higher if $\Gamma$ is lower, provided that $\Gamma$ is consistent with synchrotron/SSC model fits and minimum Lorentz factor estimates inferred from, e.g.,  $\gamma\gamma$ opacity arguments. 

Roughly, half of the pions produced by photomeson production are charged, and each neutrino carries $\sim 1/4$ of the pion energy, so the total (isotropic-equivalent) neutrino luminosity at given $E_\nu \approx 0.05 E_p$ is estimated to be $E L_E^{\nu} \sim (3/8) f_{p \gamma}  (E_p) E L_E^{\rm CR} \simeq 3.8 \times {10}^{42}~{\rm erg}~{\rm s}^{-1}~f_{p \gamma,-2} (E L_E^{\rm CR}/ {10}^{45}~{\rm erg}~{\rm s}^{-1})$ for a source satisfying $f_{p \gamma} \lesssim 1$ like BL Lac objects, where $f_{p \gamma}={10}^{-2} f_{p \gamma,-2}$ is used.  Then, one finds that the neutrino flux from an individual source is typically too low to be detected with IceCube.  One can also see that the cumulative background flux from high-peaked BL Lac objects is low.  
The UHECR energy input in the local universe is $\sim 5 \times {10}^{43}~{\rm erg}~{\rm Mpc}^{-3}~{\rm yr}^{-1}$ at ${10}^{19}$~eV~\cite{mt09}, so that assuming that such BL Lac objects and FR-I galaxies are the main UHECR sources, the expected cumulative muon neutrino background flux is estimated to be~\cite{mb10}
\begin{equation}
E_{\nu}^2 \Phi_{\nu} \sim {10}^{-10}~{\rm GeV}~{\rm cm}^{-2}~{\rm s}^{-1}~{\rm sr}^{-1}~\left( \frac{f_{p \gamma} (20 E_\nu)}{{10}^{-2}} \right) E_{\nu,17.7}^{2-p} f_z,  
\end{equation}
where $p$ is the cosmic-ray spectral index and $f_z$ is a pre-factor coming from the redshift evolution of the sources.  Low photomeson production efficiencies also follow if the UHECRs are heavy nuclei, whose losses are dominated by photodisintegration (see below).  
More luminous blazars, including low-peaked BL Lac objects, may however lead to higher photomeson production efficiencies, so that the cumulative neutrino background could be dominated by this class of AGN~\cite{muc03}. 

Our evaluation is based on the standard synchrotron/SSC model for BL Lac objects and FR-I radio galaxies.  One could abandon the standard synchrotron/SSC model and consider a highly magnetized, $\sim 10-100$~G jet model, which is needed in hadronic blazar models to accelerate protons to $\gtrsim {10}^{20}$~eV~\cite{aha00}.  
Correspondingly, the minimum magnetic luminosity is estimated to be $L_B \approx 2 \times {10}^{47}~{\rm erg}~{\rm s}^{-1}~\Gamma_1^2 {(E_{p,20}^{\rm max})}^2$, which is larger than the typical synchrotron luminosity of BL Lac objects, $L_{\gamma}^s \sim {10}^{46}~{\rm erg}~{\rm s}^{-1}$.  In the hadronic models, $\gamma$-ray emission is attributed to proton synchrotron radiation and/or proton-induced cascade emission, which leads to the requirement that the UHECR luminosity is $L_{\rm UHECR} \gtrsim L_\gamma^{C} = A_C L_{\gamma}^s$.  In the proton synchrotron blazar model~\cite{aha00,mp01,muc03}, which is typically viable for high-peaked BL Lac objects, proton synchrotron radiation is emitted up to energies $\varepsilon_{s}^{M} \approx 4 \delta_1/(1+z)$~TeV (in the limit that the maximum energy is determined by the synchrotron cooling) in efficient Fermi acceleration scenarios.    
The photomeson production efficiency for protons is strongly dependent on $\delta$, but the condition $L_{B} \gg L_{\gamma}^s$ suggests that the synchrotron energy-loss is dominant at ultra-high energies where the proton synchrotron radiation is typically prominent at $\sim$~TeV energies (though the photohadronic cascade component may become relevant at lower energies).  The strong magnetic field also suppresses electronic SSC emission because fewer electrons are needed to generate the same synchrotron flux.  

\subsection{Survival of Nuclei in the Source}

If the standard synchrotron/SSC scenario holds for TeV blazars and their misaligned counterparts, then the protons can hardly reach ${10}^{20}$~eV, as shown in Table 2 (see $E_A^{\rm max}$ obtained by detailed modeling in the literature).  
For BL Lac objects and FR-I galaxies to be the steady sources of UHECRs, therefore, UHECRs would primarily be heavier nuclei.  In such a scenario, one has to examine whether ions can survive photodisintegration losses~\citep[cf.][]{mur08a,wrm08,pmm09}.  The photodisintegration opacity is estimated similarly to the photomeson production efficiency.  
Approximating the photodisintegration cross section by the giant dipole resonance (GDR) cross section as $\sigma_{A \gamma} \sim \sigma_{\rm GDR} \delta(\varepsilon-\bar{\varepsilon}_{\rm GDR}) \Delta \bar{\varepsilon}_{\rm GDR}$, for a sufficiently soft photon spectrum (with $\alpha \gtrsim 1$), we get~\citep[][see also Murase et al. 2008a for the non-GDR effect]{mb10} 
\begin{equation}
\tau_{A \gamma} \approx \frac{t_{\rm var}'}{t_{A \gamma}'} \simeq \frac{2 \sigma_{\rm GDR}}{1+\alpha} \frac{\Delta \bar{\varepsilon}_{\rm GDR}}{\bar{\varepsilon}_{\rm GDR}} 
\frac{L_{\gamma}^s}{4 \pi \delta^4 t_{\rm var} c^2 \varepsilon^s} {\left(\frac{E_A}{E_A^b} \right)}^{\alpha-1}, 
\end{equation}
where $t_{A \gamma}'$ is the photodisintegration interaction time, $\sigma_{\rm GDR} \approx 1.45 \times {10}^{-27} A~{\rm cm}^2$, $\bar{\varepsilon}_{\rm GDR} \approx 42. 65 A^{-0.21}$~MeV (for $A>4$), $\Delta \bar{\varepsilon}_{\rm GDR} \sim 8$~MeV, and $E_A^b \approx 0.5 \delta^2 (m_A c^2 \bar{\varepsilon}_{\rm GDR}/\varepsilon^s) {(1+z)}^{-2}$ (in the observer frame).  Then we numerically find 
\begin{equation}
\tau_{A \gamma} (E_A) \simeq 0.16~\left(\frac{2.5}{1+\alpha} \right) L_{\gamma,46}^s t_{\rm var,4}^{-1} \delta_1^{-4} {\left( \frac{\varepsilon_s}{1~\rm keV} \right)}^{-1} {\left( \frac{E_A}{E_A^b} \right)}^{\alpha-1},
\end{equation}
where $E_A^b \simeq 4.8 \times {10}^{16}~{\rm eV}~{(A/56)}^{0.79} {(\varepsilon_s/1~\rm keV)}^{-1} \delta_1^2 {(1+z)}^{-2}$ is the energy of a nucleus that typically interacts with a photon with $\varepsilon_s$.  
Hence, heavy nuclei with $E_A \sim (Z/26) {10}^{20.5}$~eV (given in the observer frame) undergo  some photodisintegration reactions unless $\delta$ is high enough.  The nucleus survival condition $\tau_{A \gamma} (E_A) \lesssim 1$ gives $\delta \gtrsim 17 {(Z/26)}^{0.1} {(A/56)}^{-0.079} {(L_{\gamma,46}^s)}^{1/5} t_{\rm var,4}^{-1/5} {(\varepsilon_s/1~\rm keV)}^{-0.1} {(1+z)}^{-1/5}$ (for $\alpha \sim 1.5$), but significant photodisintegration loss is easily avoided for reasonably large bulk outflow Doppler factors.  

Recalling from equation~(\ref{fpgamma}) that the photomeson production efficiency has the same dependence on $\delta$, we can conclude that when heavy nuclei survive photodisintegration, the photomeson production efficiency is so low that the corresponding neutrino and $\gamma$-ray fluxes are not easily detected~\citep[][but see also Murase \& Beacom 2010b]{mb10}. 

\section{Extreme TeV Blazars and Intergalactic Cascades}

The $\gamma\gamma$ opacity argument allows us to place constraints on $\delta$ in BL Lac objects observed at TeV energies, and requires $\delta \gtrsim 60$ for PKS 2155-304~\citep{bfr08} for the major July/August 2006 TeV flares~\citep{aha07a}, and $\delta \gtrsim 100$ to be furthermore consistent with synchrotron/SSC model fitting for different models of the extragalactic background light~\citep[EBL;][]{fdb08}.  
Another important fact is that VHE photons can interact with the cosmic photon backgrounds.  VHE $\gamma$ rays produce electron-positron pairs via $\gamma \gamma$ pair creation, and the resulting high-energy pairs make high-energy photons via Compton scattering.  Hence, the cascaded $\gamma$ rays, which are often called pair echoes~\citep[e.g.,][]{pla95,mur08b} and/or pair haloes~\citep[e.g.,][]{acv94,ns07}, are expected at GeV-TeV energies.  In particular, $\gamma$ rays with energies below $\sim 100$~TeV are likely to leave structured regions of the universe, inducing the cascade in the void region~\cite{mur08b}.  

Many $\gamma$-ray blazars show variability, and often display spectacular flares.  Some of them are ultra-variable, as seen in multi-TeV flaring episodes from PKS 2155-304~\cite{aha07a,aha09b}, Mrk 501~\cite{alb07b} and Mrk 421~\cite{gal11}.   Such rapidly varying $\gamma$-ray emission should be produced in the blazar region.  This is because the IGMF will introduce a significant time spread in the cascade radiation, which seems incompatible with rapidly varying emissions.  
For example, consider primary 10 TeV $\gamma$ rays, so the Compton-upscattered cosmic microwave background (CMB) photons have $E_{\gamma} \approx (4/3) \gamma'^2 \varepsilon_{\rm CMB} \simeq 88~{\rm GeV}~{\gamma'}_7^2$.  Based on the lower limits of $B_{\rm IGV} \lambda_{\rm coh}^{1/2} \gtrsim 10^{-18} - 10^{-17}~{\rm G}~{\rm Mpc}^{1/2}$~\cite{dol11,der11,tak11}  obtained for 1ES 0229+200 (where the term $\lambda_{\rm coh}$ is the coherence length of the magnetic field, and this relation is understood to apply when $\lambda_{\rm coh}$ is smaller than the cooling length for GeV production), the timescale of the $\sim 0.1$~TeV pair echo is estimated to be~\cite{mur08b,der11},
\begin{equation} 
{\Delta t}_{\rm IGV} \simeq 1.4~{\rm yr}~E_{\gamma,11}^{-2} B_{\rm IGV,-17}^2 {(\lambda_{\gamma \gamma}/100~\rm Mpc)} {(1+z)}^{-1}, 
\end{equation}
where $\lambda_{\gamma \gamma}$ is the mean free path for the $\gamma \gamma$ pair creation, and the void IGMF $B_{\rm IGV}$ is defined in the frame of the Hubble flow.  
Therefore, the highly variable VHE radiation from these BL Lac objects is likely to be either leptonic synchrotron or SSC, or proton synchrotron radiation produced in the jet (even though another slowly variable component may be produced by the secondary emission).

This conclusion does not however hold for a fraction of blazars and radio galaxies from which prominent variability has not been seen.  An interesting source is the extreme TeV blazar 1ES 0229+200, which has a hard VHE component extending to $> 10$ TeV, but has not been reported to be variable in observations taken over a period of $3-4$~yr~\citep{aha07b,per10}.  
If the apparent absence of the variability comes from observational limitations and if fast variability is present, the emission should be produced in/near the blazar region.  If it is the case that there is no rapid variability, the observed component may come from an extended jet~\citep[][see also Section 4 for further discussion on the $\gamma$-ray emission region]{bdf08}.  In addition, as we see in this paper, a slowly variable component can be $\gamma$-ray-induced intergalactic cascade emission.  Furthermore, if it is non-variable, proton-induced intergalactic cascade emission (i.e., intergalactic cascades caused by UHE $\gamma$ rays and pairs generated via the photomeson production with the CMB and EBL) can be responsible for the observed emission.  These cascade emissions may confuse interpretation of the minimum bulk Lorentz factor from $\gamma \gamma$ opacity arguments and the level of the EBL~\cite{ek10,ess10,ess11}.  The intergalactic cascade components could be present not only in extreme blazars, but also on longer timescales in other blazars and radio galaxies.  For example, a slowly variable ($\sim$~month) emission at GeV energies was observed from Mrk 501~\cite{abd11c}, which could arise from the intergalactic cascade induced by variable TeV source photons~\cite{nst11}.  

Here we focus on the intergalactic cascade scenarios in order to explain hard VHE spectra of extreme TeV blazars whose variability is apparently absent.  We calculate the cascade emission by solving the Boltzmann equations, where $\gamma \gamma$ pair creation, IC scattering, synchrotron radiation, and adiabatic energy loss are taken into account~\cite{lee98}.  As for proton propagation, we directly solve the equation of motion of protons one by one, with photomeson production simulated by SOPHIA~\cite{muc00} and the Bethe-Heitler process included to treat interactions with the ambient photon field~\cite{cho92}.  Then the electromagnetic cascade is calculated separately.  For the EBL model, we employ the low-IR and best-fit models~\citep[][see Finke et al. 2011 for detailed discussions on the EBL]{kne04,kd10}.  We focus on the possibility that the cascade interpretation is a viable explanation of VHE $\gamma$-ray spectra of extreme TeV blazars.  Here, the IGMF in voids has to be weak enough ($B_{\rm IGV} \lambda_{\rm coh}^{1/2} \lesssim {10}^{-15}~{\rm G}~{\rm Mpc}^{1/2}$) that the cascade radiation at TeV energies is not suppressed by the IGMF if it is to make the measured flux in the VHE range.  On the other hand,  the void IGMF cannot be below $\sim {10}^{-18}~{\rm G}~{\rm Mpc}^{1/2}$ due to constraints from \textit{Fermi}~\citep[e.g.,][]{dol11,der11,tav11,tay11,as11}.

An important point is that cosmic magnetic fields are almost certainly inhomogeneous.  Whereas one may expect very weak IGMFs in voids, $B_{\rm IGV} \lambda_{\rm coh}^{1/2} \ll {10}^{-9}~{\rm G}~{\rm Mpc}^{1/2}$, the structured region of the universe is likely to be significantly magnetized.  Clusters of galaxies are known to have $B_{\rm EG} \sim 0.1-1~\mu {\rm G}$~\citep[e.g.,][]{val04}, and recent simulations have suggested that filaments have $B_{\rm EG} \sim 1-10~{\rm nG}$~\citep[][and see also, e.g., Donnert et al. 2009]{ryu08,das08}, which are larger than levels expected for the IGMF in voids, and galaxies including AGN are likely located in these structured regions of the universe.  The mean free path of $\lesssim 100$~TeV and $\gtrsim 3$~EeV $\gamma$ rays is larger than $\sim$~Mpc~\citep[e.g.,][]{der07}, so that one may expect that the cascade emission induced by VHE/UHE primary $\gamma$ rays is primarily developed in the voids.  On the other hand, ions must propagate in the clusters and/or filaments, so that they are deflected (and delayed) by their magnetic fields.  Indeed, as demonstrated by a number of authors, the structured EGMFs play a crucial role on propagation of UHECRs~\citep[e.g.,][]{tys06,das08}, and this is even more so the case for lower-energy cosmic rays.    

\subsection{Cascades by Primary VHE/UHE Gamma Rays}

SEDs of high-peaked BL Lac objects are generally well reproduced by the standard one-zone electronic synchrotron/SSC model.  Among them, extreme TeV blazars have the hardest VHE $\gamma$-ray spectra at $\sim 1-10$~TeV energies, as indicated by deabsorption of the measured $\gamma$-ray spectrum based on conventional EBL models (discussed below) and supported by non-detections of GeV $\gamma$ rays by \textit{Fermi}.  Also, in some cases (RGB J0152+017, 1ES 0229+200 and 1ES 0548-322), the optical/UV data show a rather steep spectrum which is thought to be the emission from the host galaxies~\cite{tav11}.  
Although the synchrotron component of extreme TeV blazars seem unremarkable at the optical/UV band, in these cases, comparison between optical/UV and X-ray data requires a strong roll-off of the nonthermal spectrum below the X-ray band, suggesting that $F_\nu^s \propto \nu^{1/3}$ for 1ES 0229+200~\cite{tav11}. 

It is possible to explain such hard $\gamma$-ray spectra by the SSC model, but extreme parameters seem necessary compared to cases of typical, variable high-peaked BL Lac objects.  It often suggests a very narrow-range energy distribution of electrons, and unusually large values of $\delta \sim {10}^2 - {10}^3$ may be necessary to avoid the Klein-Nishina suppression.  
For 1ES 0229+200, PKS 0548-322 and 1ES 0347-121, extreme values of the electron minimum Lorentz factor of $\gamma_{e,m} \sim {10}^{4}-{10}^{5}$ are required from spectral modeling, where the hard SSC spectrum, $F_E \propto E^{-2/3}$ (in the Thomson regime) can be expected in the VHE range~\cite{tav11}.  

There are several alternate blazar models that predict very hard $\sim 30-100$~TeV $\gamma$-ray emission.  In the hadronic model, the proton synchrotron process leads to multi-TeV emission if the outflow is ultra-relativistic, $\Gamma \sim {10}^2-{10}^3$.  Then, further hardening may be caused by internal absorption due to some soft photon field outside the blob~\cite{zac11}. Another possibility to make hard TeV emission is electromagnetic radiation produced by nonthermal electrons in the vacuum gap of the black hole magnetosphere~\citep[e.g.,][]{lev00}.  

B\"ottcher et al. (2008) suggested that hard VHE emission originates from CMB photons Compton-upscattered by relativistic electrons that are accelerated in the extended jet.   In this model, if the electron spectrum is hard, $p \sim 1.5$, the resulting number spectrum of the IC emission has $F_E \propto E^{-(1+p)/2} \sim E^{-5/4}$, which is compatible with the observed VHE $\gamma$-ray spectrum.  The same process could be important for recollimation shocks \citep[e.g.,][]{bl09,nw00}, or acceleration at knots and hotspots, noting that variations on much longer timescales can be expected in these models. 

\begin{figure}[tb]
\includegraphics[width=1.00\linewidth]{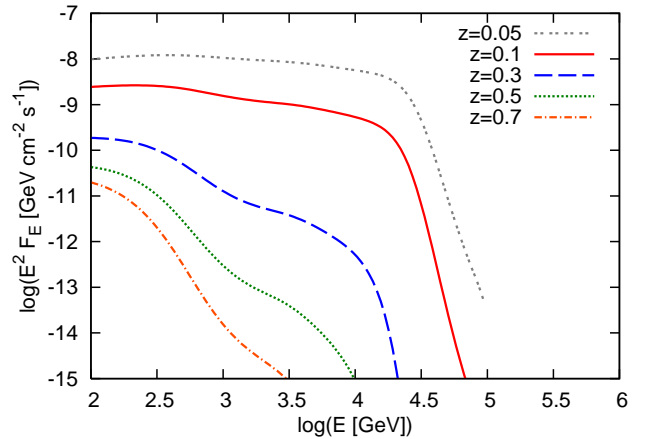}
\caption{Spectra of VHE $\gamma$-ray-induced cascade emission for various source redshifts.  We assume the total $\gamma$-ray luminosity of $L_{\gamma}={10}^{45}~{\rm erg}~{\rm s}^{-1}$ with $\beta=2/3$ and $E^{\rm max}=100$~TeV.  The low-IR EBL model of Kneiske et al.\ (2004) is used here.}
\label{fig1}
\end{figure}

\begin{figure}[tb]
\includegraphics[width=1.00\linewidth]{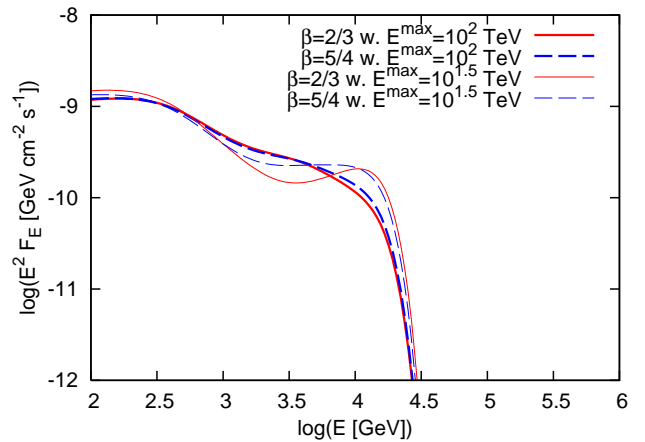}
\caption{Spectra of VHE $\gamma$-ray-induced cascade emission for various intrinsic photon spectra.  The source redshift is set to $z=0.14$.}
\label{fig2}
\end{figure}

When VHE $\gamma$ rays are emitted from a source, they induce an electromagnetic cascade in intergalactic space.  This cascade unavoidably accompanies spectral production of extreme TeV blazars as long as the IGMF in voids is weak enough.  To demonstrate this,  we show in Figures~\ref{fig1} and \ref{fig2} the VHE $\gamma$-ray-induced cascade emission for sources at various redshifts.  In Figure~\ref{fig1}, primary source photons with $F_E \propto E^{-\beta}$ with $\beta=2/3$ and $E^{\rm max} =100$~TeV (in the source rest frame) are assumed.  One sees that the observed cutoff due to the EBL becomes lower for more distant sources since the $\gamma \gamma$ pair-creation opacity increases.  In Figure~\ref{fig2}, a different photon index ($\beta=5/4$) and/or a different maximum energy ($E^{\rm max}={10}^{1.5}$~TeV) are assumed for comparison, which causes slight differences in spectra.  As indicated by equation~(9), the intergalactic cascade emission induced by primary $\gamma$ rays will be slowly variable or almost steady.


Another way to have $\gamma$-ray induced emission involves UHE $\gamma$ rays produced in blazar jets or radio galaxies.  Such a case is shown in Figure~3 assuming much higher injected photon energies than before. Here we assume an injection spectrum centered at 10~EeV spanning one decade.  The photomeson production by UHE protons, which can be expected in hadronic models, leads to UHE photons with energy $E_\gamma \approx 0.1 E_p \simeq {10}^{19}~{\rm eV}~E_{p,20}$.  In the synchrotron source in which UHE protons are accelerated, one may expect that the synchrotron self-absorption cutoff curtails the number of low energy photons impeding UHE photon escape from the emission region~\cite{mur09}.  A caveat of this model in our case is that generation of UHE $\gamma$ rays in the source requires acceleration of UHE protons and moderately efficient photomeson production.  As noted before, the photomeson production in the source may not be too efficient in high-peaked BL Lac objects, which implies that the required UHECR luminosity has to be very large.  As can be seen, there is some notable differences at low redshifts $z \lesssim 0.1$ due to the longer effective energy loss length of UHE $\gamma$ rays, but the received spectra are not strongly sensitive to the energy at which the photons are injected for higher redshift sources.

\begin{figure}[tb]
\includegraphics[width=1.00\linewidth]{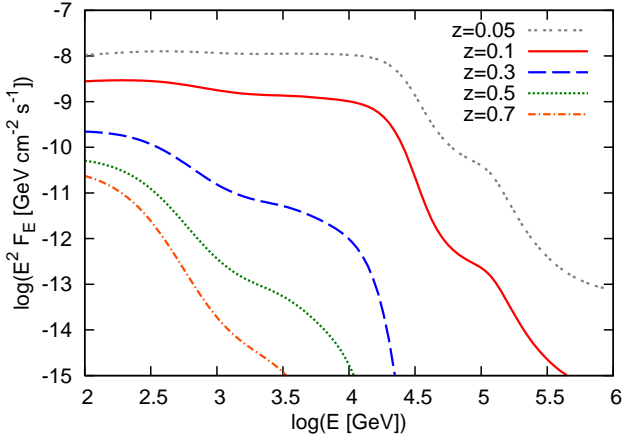}
\caption{Spectra of UHE $\gamma$-ray-induced cascade emission for various source redshifts. We assume $L_{\gamma}={10}^{45}~{\rm erg}~{\rm s}^{-1}$ at ${10}^{18.75}-{10}^{19.25}$~eV.}
\label{fig3}
\end{figure}

Note that the intergalactic cascade scenario makes a non-variable or slowly variable component, even when the $\gamma$-ray emission made in the jet contributes to a separate highly variable component.  Although there is no strong evidence of time variability for several extreme TeV blazars, future sensitive observations by CTA \citep{cta10}, HAWC \cite{san09}, LHAASO~\citep{cao10}, or SCORE~\citep{hth11} will be crucial for identifying a slowly variable $\gamma$-ray emission component.

\subsection{Cascades by Primary UHECRs}

In the previous subsection, we considered cascade emission induced by primary $\gamma$ rays.  VHE $\gamma$ rays at $\lesssim 100$~TeV, or UHE $\gamma$ rays with energies $\gtrsim 3$~EeV,  where the opacity of the background radiation is not so large, can leave structured regions of the universe, whereas cosmic rays should feel structured EGMFs in clusters and filaments.  The deflection of cosmic rays by the structured region with size $l \sim$~Mpc,  magnetic field of $B_{\rm EG} \sim 10$~nG, and coherence length of $\lambda_{\rm coh} \sim 0.1$~Mpc (which may be typical of filaments; Ryu et al. 2008) is estimated to be~\cite{tm11}
\begin{equation}
\theta_{\rm CR} \approx \frac{\sqrt{2 \lambda_{\rm coh} l}}{3 r_L} \simeq 8^{\circ}~Z E_{A,19}^{-1} B_{\rm EG,-8} {\left( \frac{\lambda_{\rm coh}}{0.1~\rm Mpc} \right)}^{1/2} {\left( \frac{l}{\rm Mpc} \right)}^{1/2}.
\end{equation}
Therefore, the deflection by the structured EGMFs is not negligible for cosmic rays with energies $\lesssim {10}^{19}$~eV, since the deflection angle is larger than the typical jet opening angle of $\theta_j \sim 0.1 \sim 6^\circ$.  The corresponding time spread due to a structured EGMF around the source (that is comparable to the time delay) is expected to be
\begin{eqnarray}
\frac{\Delta t_{\rm CR}}{1+z} &\approx& \frac{1}{4} \theta_{\rm CR}^2 \frac{l}{c} \nonumber \\
&\simeq& 2 \times {10}^{4}~{\rm yr}~Z^2 E_{A,19}^{-2} B_{\rm EG,-8}^2 {\left( \frac{\lambda_{\rm coh}}{0.1~\rm Mpc} \right)} {\left( \frac{l}{\rm Mpc} \right)}^{2},
\end{eqnarray} 
which is unavoidable as long as cosmic rays pass through the structured region around the source, and it implies that the resulting cascade emission is essentially regarded as steady emission.  Notice that the total time spread ${\Delta T}_{\rm CR}$ could generally be longer than ${\Delta t}_{\rm CR}$ due to additional time spread by intervening structured EGMFs and the void IGMF.  

In order to model the structured EGMFs, we have assumed a simplified two-zone model with structured EGMF and IGMF in voids~\citep[see][for details]{tm11}.
We model a cluster of galaxies by a sphere with the radius of 3 Mpc, and $B_{\rm EG} (r)=B_0 {(1+r/r_c)}^{-0.7}$, with $B_0 = 1~\mu \rm G$ and $r_c=378$~kpc.  The magnetic field direction is assumed to be turbulent with the Kolmogorov spectrum and the maximum length of $\lambda_{\rm max}=100$~kpc.  In addition to the EBL, the infrared background in the cluster is considered as the superposition of the SEDs of 100 giant elliptical galaxies calculated by GRASIL~\cite{sil98}, using fitting formula for the gas distribution~\cite{rgd04}.  Filaments are modeled by a cylinder with a radius of $2$~Mpc~\cite{ryu08} and a height of $25$~Mpc.  
The magnetic field is assumed to be turbulent, which is described by the Kolmogorov spectrum with $B_{\rm EG} = 10~\rm nG$ and $\lambda_{\rm max}=100$~kpc, although these values are very uncertain.  Some numerical simulations imply a large-scale coherent component of the magnetic field in filaments~\citep[e.g.,][]{bru05}, which may deflect cosmic-ray trajectories even more effectively.  UHECRs are injected from the center of the filament toward a direction perpendicular to the cylindrical axis in order to examine a relatively conservative case.  Throughout this work, the IGMF in voids is assumed to be weak enough to be less important for cosmic-ray deflections.  

\begin{figure}[tb]
\includegraphics[width=1.00\linewidth]{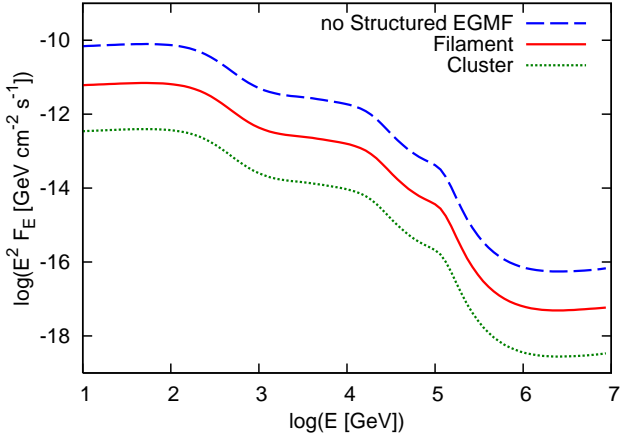}
\caption{Effects of the structured EGMF on the $\gamma$-ray flux. We assume $L_{\rm UHECR}={10}^{45}~{\rm erg}~{\rm s}^{-1}$, with $E_{p}^{\rm max}={10}^{19}$~eV and $p=2$.  Here, as in the results on cascade emission induced by primary $\gamma$ rays, we use the isotropic-equivalent cosmic-ray luminosity at the source (defined for UHECRs above ${10}^{18.5}$~eV), which is related to the absolute (beaming-corrected) cosmic-ray luminosity, $L_{{\rm UHECR},j}$, as $L_{\rm UHECR} \equiv (1-\cos \theta_j)^{-1} L_{{\rm UHECR},j}$.  Here the assumed jet opening angle is $\theta_j=0.1$.  The source redshift is set to $z=0.5$.}
\label{fig4}
\end{figure}

\begin{figure}[tb]
\includegraphics[width=1.00\linewidth]{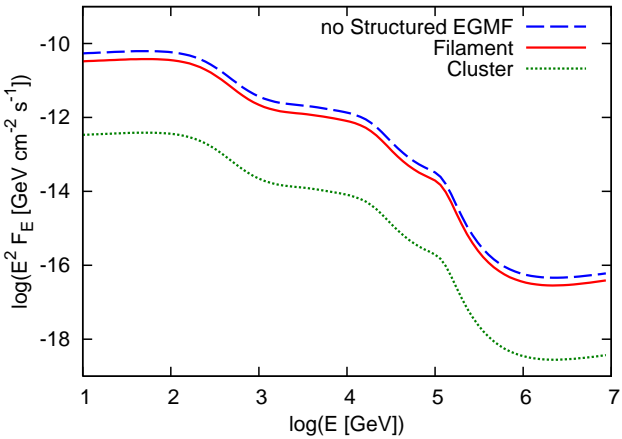}
\caption{Same as Figure~4, but with $E_{p}^{\rm max}={10}^{20}$~eV.}
\label{fig5}
\end{figure}

In Figure~\ref{fig4}, we show our numerical results for the case $E_{p}^{\rm max} = {10}^{19}$~eV expected in the standard synchrotron/SSC model of typical, variable BL Lac objects and FR-I galaxies.  One sees that the structured EGMFs play an important role by suppressing the resulting $\gamma$-ray flux by more than one order of magnitude compared to the case without them.  In Figure~\ref{fig5}, we show the case of $E_{p}^{\rm max} = {10}^{20}$~eV, which can be achieved in the hadronic model.  While the Bethe-Heitler pair-creation process dominantly provides an electromagnetic component in Figure~\ref{fig4}, contribution of photomeson production is more important in Figure~5.  In the filament case, the deflection angle of UHECRs around ${10}^{20}$~eV is still less than the jet opening angle, so that the $\gamma$-ray flux is diluted by only a small factor.  On the other hand, in the cluster case, because UHECRs cannot be beamed, the $\gamma$-ray flux becomes almost isotropic and the corresponding flux is reduced according to the jet beaming factor $(1-\cos \theta_j) \simeq 1/200$ for $\theta_j = 0.1$.  The effects of the structured EGMFs are illustrated in Figure~6, where the relative contributions are calculated from two-dimensional Gaussian fits.  Note that if we express the isotropic-equivalent cosmic-ray luminosity where cosmic rays leave the structured region as $E L_E^{\rm CR}$, then the relative contributions are $(1-\cos \theta_j) (E L_{E}^{\rm CR})/(E L_{E}^{{\rm CR},j})$.  In the filament case, isotropization becomes significant at $\sim {10}^{19}$~eV rather than at $\sim {10}^{21}$~eV for the cluster case.      
   
\begin{figure}[tb]
\includegraphics[width=1.00\linewidth]{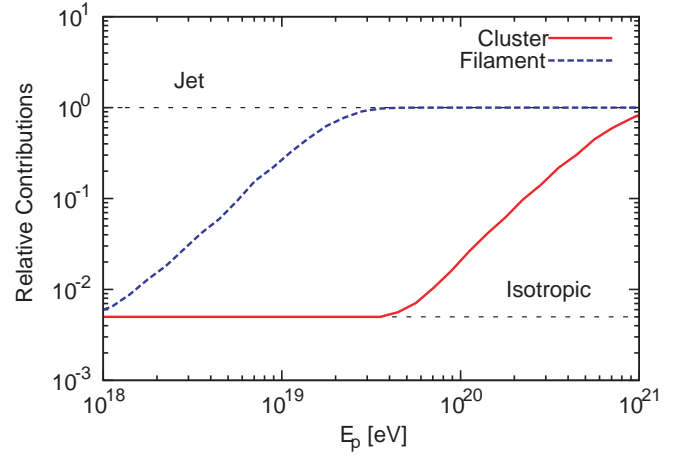}
\caption{Effects of the structured EGMFs on the deflection of UHE protons.  Relative contributions represent how much the apparent cosmic-luminosity at which cosmic rays enter the void region is diluted from $E L_E^{\rm CR}$ at the source.  Note that a two-sided jet is considered throughout this work.
}
\label{fig6}
\end{figure}

\begin{figure}[tb]
\includegraphics[width=1.00\linewidth]{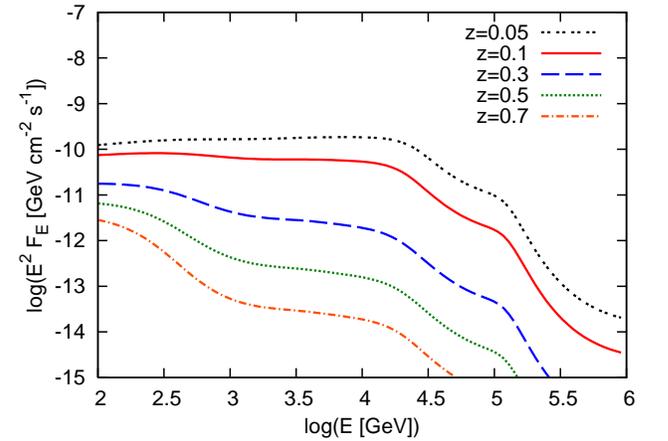}
\caption{Spectra of UHE proton-induced cascade emission for various source redshifts.  We assume $L_{\rm UHECR}={10}^{45}~{\rm erg}~{\rm s}^{-1}$ with $E_{p}^{\rm max}={10}^{19}$~eV and $p=2$.  The source is assumed to be located in the filament with $B_{\rm EG} = 10$~nG and $\lambda_{\rm max} = 0.1$~Mpc.  The low-IR EBL model is here assumed.}
\label{fig7}
\end{figure}

In Figure~\ref{fig7}, we show resulting $\gamma$-ray spectra for various redshifts. Owing to the Bethe-Heitler process with energy-loss length $\sim$~Gpc, UHE protons continue to supply electron-positron pairs for a longer distance than the photomeson energy loss length of $\sim 100$~Mpc.  As a result, the dependence of the proton-induced $\gamma$-ray fluxes on distance is much gentler than $\gamma$-ray-induced fluxes.  Indeed, one sees that the relative importance of the proton-induced $\gamma$-ray flux to the $\gamma$-ray induced flux increases with distance (compare Figure~7 with Figures~1 and 3).  Importantly for distant sources, the proton-induced cascade spectrum is much harder than the $\gamma$-ray induced spectrum, especially above TeV energies. Future VHE observations by CTA and HAWC are important to identify the origin of UHECRs through detection of high-energy $\gamma$ rays, as we now demonstrate for 1ES 0229+200 in the next subsection. 
 
In this work, we are interested in cases where IC cascade emission in voids is important in the VHE range, since it can explain hard VHE spectra of extreme TeV blazars as suggested by Essey et al. (2010).  When pairs are mainly supplied via the Bethe-Heitler process, the timescale of secondary photons produced by a proton beam roughly becomes
\begin{equation} 
{\Delta t}_{\rm IGV} \simeq 14~{\rm yr}~E_{\gamma,11}^{-2} B_{\rm IGV,-17}^2 {(\lambda_{\rm BH}/\rm Gpc)} {(1+z)}^{-1}, 
\end{equation}
which is more relevant than ${\Delta T}_{\rm CR}$ when the void IGMF is so strong that ${\Delta T}_{\rm CR} < {\Delta t}_{\rm IGV}$ is satisfied. Here, $\lambda_{\rm BH}$ is the Bethe-Heitler energy loss length. 
One should also keep in mind that the proton-induced GeV-TeV synchrotron emission from the structured region itself, where the EGMFs are stronger, should also be expected~\citep[see][and references therein]{ga05,kot09,kot11}.  For a weak IGMF that is of interest in this work, its relative importance is somewhat smaller when the volume filling fraction of the magnetized region is taken into account.   

We have demonstrated the likely importance of the structured EGMFs for proton-induced intergalactic cascade emission.  They are also important for UHE nuclei.  Since nuclei with energy $Z E_p$ have the same deflection angle as protons with energy $E_p$, our results indicate that Fe nuclei should be significantly isotropized for all observed UHECR energies.  For UHE nuclei, the photodisintegration energy loss length is $\sim 100$~Mpc, for which the energy fraction carried by $\gamma$ rays and neutrinos is small as long as $E_A^{\rm max}$ is not too high.  On the other hand, UHE nuclei supply high-energy pairs via the Bethe-Heitler process, whose effective cross section is $\kappa_{{\rm BH},A} \sigma_{{\rm BH},A}  \sim \kappa_{{\rm BH},p}  \sigma_{{\rm BH},p} (Z^2/A)$, which induces cascades in the same manner as UHE protons.  Therefore, \textit{the intergalactic cascade signal, which is generated outside the source, is also important for sources of primary UHE nuclei}.~\footnote{On the other hand, the emission of $\gamma$ rays and neutrinos produced inside the source of primary UHE nuclei is limited by the nuclear survival condition, as shown in Murase \& Beacom (2010a; 2010b).  Given that the observed UHECRs are dominated by heavy nuclei, this limitation is also applied to neutrinos produced outside the source, i.e., cosmogenic neutrinos~\cite{mb10}.}

\subsection{Implications for TeV-PeV Observations}

In a wide range of EBL models, deabsorption of measured TeV blazar spectra leads to hard excesses at $>$TeV energies in, e.g., 1ES 1101-232, 1ES 0229+200, and 1ES 0347-121~\citep[see, e.g., Fig.\ 8 in][]{frd10}.  
 These unusual TeV spectral emission components are conventionally explained by (either leptonic or hadronic) emissions at the source, but they could also be explained by intergalactic cascade emissions.
Non-simultaneous TeV excesses are also seen above the extrapolation of the GeV flux in NGC 1275~\citep{abd09ngc1275} and the core of Cen A~\citep{abd10cena}, but because of their proximity, these excesses are unlikely to be UHECR-induced emissions made in intergalactic space.

\begin{figure}[tb]
\includegraphics[width=1.00\linewidth]{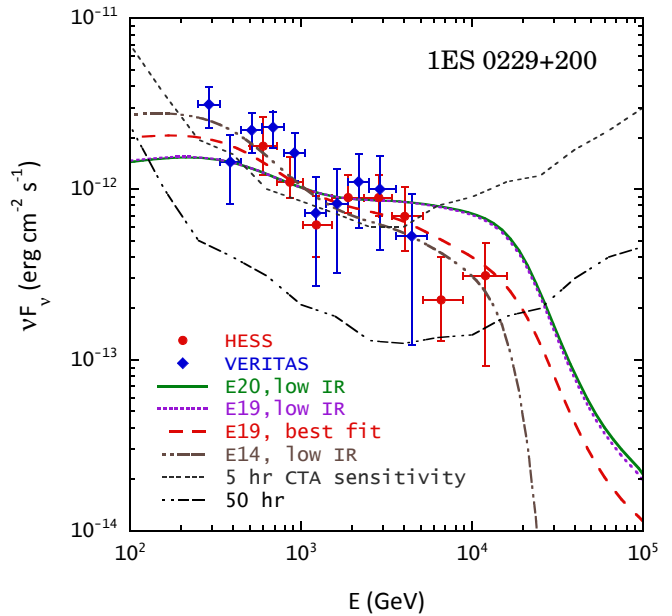}
\caption{Spectral fits to HESS and VERITAS data of 1ES 0229+200.  Blue data points are from HESS~\citep{aha07b}, and red data points are preliminary VERITAS data~\cite{per10}.  The curves labeled ``E20, low IR" and ``E19, low IR" are the cascade spectra initiated by the $E^{-2}$ injection with $E_p^{\rm max} = {10}^{20}$ eV and ${10}^{19}$ eV protons, respectively, using the low-IR EBL model \citep{kne04}, whereas the curve labeled ``E19, best fit" is the spectrum with $E_p^{\rm max} = 10^{19}$ eV for the best-fit EBL model.  The curve labeled ``E14, low IR" is the spectrum resulting from the cascade of $E^{\rm max}={10}^{14}$ eV photons with $\beta=5/4$ produced at the source for the low-IR EBL model.  Double dot-dashed and dotted curves give, respectively, the 5$\sigma$ differential sensitivity for 5 and 50 hr observations with CTA~\citep[configuration E;][]{cta10}.  }
\label{fig8}
\end{figure}

Figure~\ref{fig8} demonstrates that 1ES 0229+200 can be fit by both the $\gamma$-ray induced cascade and proton-induced cascade emissions.  Because of the uncertainty in EBL models, it is not easy to distinguish between the two possibilities at $\sim 0.1-1$~TeV energies.  At higher energies, however, our calculations show that UHECR-induced cascade emission becomes harder than $\gamma$-ray-induced cascade emission resulting from attenuation of hard $\gamma$-ray source photons, for a given EBL model.  
More importantly, the emission spectrum measured as a result of the injection of VHE/UHE photons at the source is strongly suppressed above $\sim 10$~TeV for a wide range of EBL models, whereas a cosmic-ray-induced cascade displays a significantly harder spectrum above this energy, and detection of $>25$ TeV $\gamma$ rays from 1ES 0229+200 is only compatible if the $\gamma$ rays are hadronic in origin.  
This is because UHE protons (and UHE nuclei) can inject high-energy pairs over the Bethe-Heitler energy loss length ($\lambda_{\rm BH} \sim (A/Z^2)$~Gpc at $E_A \sim A {10}^{19}$~eV) that is typically longer than the effective loss length of VHE/UHE photons. 
For steady, non-variable $\gamma$-ray sources, this intergalactic cascade signal induced by UHECRs provides a crucial probe of UHECR sources.  Its identification would demonstrate that a distant blazar is an UHECR source through electromagnetic channels, which provides another important clue besides $\gamma$-ray variability.  Identifying this feature by future Cherenkov detectors such as CTA or HAWC is possible, and the differential sensitivity goal of CTA is shown~\citep{cta10}.  
Note that this is a differential sensitivity curve with the requirement of 5$\sigma$ significance for 50 hour observations per bin, with 4 bins per decade.  This is a much more stringent requirement than detection of a source with 5$\sigma$ based on integrated flux, which can be divided into 3 data points with $\approx 3\sigma$ significance each.  Given the differential CTA sensitivity for a 50 hr observation, the spectral hardening associated with hadronic cascade development can be clearly detected.

It is theoretically expected that cosmic-ray-induced and $\gamma$-ray-induced cascade emissions are more easily discriminated in higher redshift sources.  For the $\gamma$-ray-induced cascade, there should be a cutoff because of $\gamma \gamma$ pair creation by the EBL, while spectra of the cosmic-ray-induced cascade emission are hardened by the continuous injection through the Bethe-Heitler process.  Hence, deep observations at $\gtrsim$~TeV energies by CTA or HAWC for moderately high-redshift blazars will also be important to resolve this question, along with detailed theoretical calculations for individual TeV blazars.   

Now that IceCube has been completed,  it has started to give important insights into the origin of UHECRs by itself and with GeV and VHE $\gamma$-ray observations.  But detection of neutrino signals produced outside the source seems difficult for high-peaked BL Lac objects and FR-I galaxies, because the point source flux sensitivity at $>10$~PeV is order of $\sim{10}^{-11}~{\rm TeV}~{\rm cm}^{-2}~{\rm s}^{-1}$~\cite{spi11,abb11}, which is typically larger than the expected neutrino fluxes, as shown in Figure~9.  
On the other hand, the cumulative (diffuse and stacked) background neutrino flux may be detectable especially for $E_{p}^{\rm max} \gtrsim {10}^{20}$~eV~\citep[c.f.][and references therein]{anc07,tak09}, which is possible in hadronic models with large magnetic fields in jets.  For $E_{p}^{\rm max}={10}^{19}$~eV, however, protons mostly interact with the EBL, and the expected flux is lower than the proton case even if nuclei can be accelerated up to $E_A = Z E_p$~\citep[e.g.,][]{anc07}.  If BL Lac objects and FR-I galaxies are the main sources of UHECRs made mainly of ions with $E_A \lesssim Z {10}^{19}$ eV, the cumulative neutrino background would be difficult for IceCube to detect.  

\begin{figure}[tb]
\includegraphics[width=1.00\linewidth]{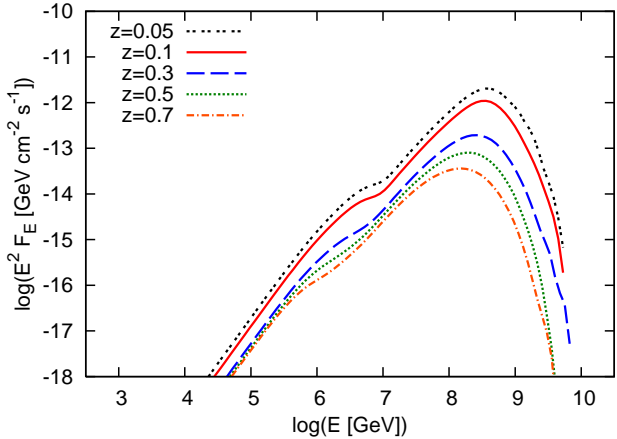}
\caption{Spectra of UHE proton-induced neutrino emission for various source redshifts.  The parameters used here are the same as Figure~7.}
\label{fig9}
\end{figure}

Next, let us discuss the UHECR luminosity required to explain such extreme TeV blazars in the intergalactic UHECR-induced cascade scenario. In Figure~8, with $E_p^{\rm max}= {10}^{19}$~eV and $p=2$, the inferred isotropic-equivalent UHECR luminosities (at the source) are $L_{\rm UHECR}\simeq {10}^{46}~{\rm erg}~{\rm s}^{-1}$ (for the filament) and $L_{\rm UHECR}\simeq 5 \times {10}^{46}~{\rm erg}~{\rm s}^{-1}$ (for the cluster), respectively, while $L_{\gamma} \simeq {10}^{45}~{\rm erg}~{\rm s}^{-1}$ when primary $\gamma$ rays are injected.  
 
If no structured EGMFs are there, the required isotropic-equivalent UHECR luminosity (at the source) is $L_{\rm UHECR} \sim {10}^{45}-{10}^{46}~{\rm erg}~{\rm s}^{-1}$ (i.e., the corresponding absolute, beaming-corrected cosmic-ray luminosity $L_{{\rm UHECR},j} \sim {10}^{43}~{\rm erg}~{\rm s}^{-1}$), which is consistent with Essey et al. (2011), Razzaque et al. (2012), and the calculation for 1ES 0229+200 shown here.  It is also consistent with the proton power needed in hadronic models for typical, variable BL Lac objects.  Where structured EGMFs play a role, the required UHECR luminosity becomes much larger.  In fact, we obtained $L_{\rm UHECR} \sim {10}^{46}-{10}^{47}~{\rm erg}~{\rm s}^{-1}$ (for filaments) and $L_{\rm UHECR} \sim {10}^{47}-{10}^{48}~{\rm erg}~{\rm s}^{-1}$ (for clusters), depending on the EBL model and spectral indices, when we assume $E_p^{\rm max}= {10}^{19}$~eV.  
Such UHECR luminosities, obtained with the structured EGMFs, seem rather extreme, since the total cosmic-ray luminosity including low-energy cosmic rays is at least $\sim 20$ times larger for $p \gtrsim 2$ and $E_p^{\rm min}=10$~GeV.  In the cluster EGMF case, this means an absolute luminosity of $L_{{\rm CR},j} \equiv \int_{E_p^{\rm min}}^{E_p^{\rm max}} d E \, L_E^{{\rm CR},j} \gtrsim {10}^{46}-{10}^{47}~{\rm erg}~{\rm s}^{-1}$, which is comparable to the Eddington luminosity of a $\sim {10}^{8}-{10}^9 M_\odot$ black hole,
\begin{equation}
L_{\rm Edd,abs} = \frac{4 \pi G M_{\rm BH} m_p c}{\sigma_T} \simeq 1.3 \times {10}^{46}~{\rm erg}~{\rm s}^{-1}~\left( \frac{M_{\rm BH}}{{10}^{8}~M_{\odot}} \right),
\end{equation} 
where $M_{\rm BH}$ is the black hole mass.  Therefore, the intergalactic UHECR-induced cascade interpretation becomes problematic, if runaway UHECRs are significantly isotropized and/or the spectral index of cosmic rays is steep enough.  Such isotropization may be realized by some plasma instability, or the structured EGMFs and/or magnetic fields in radio bubbles or lobes accompanied by radio-loud AGN (see below).    

Note that our conclusion from Table~2 does not hold in the intergalactic hadronic cascade interpretation of extreme blazars since the SSC model is here abandoned.  But one may adopt $E_p^{\rm max} \sim {10}^{19}$~eV, motivated by results of the SSC modeling for typical, variable blazars (see Table~2).  On the other hand, higher $E_p^{\rm max}$ is also possible and a proton spectrum with higher $E_p^{\rm max}$ is favored in view of smaller deflections and relaxed luminosity requirement to fit TeV data (see Figure~6).  However, similarly to the proton synchrotron blazar model for variable BL Lac objects, the proton synchrotron component is expected as well as the intergalactic hadronic cascade component.       

It is useful to compare those luminosities with the required UHECR energy budget indicated from UHECR observations.  From recent PAO observations, the local UHECR energy budget above ${10}^{18.5}$~eV is a few~$\times {10}^{44}~{\rm erg}~{\rm Mpc}^{-3}~{\rm yr}^{-1}$.  For the local blazar density, $n_s \sim {10}^{-6.5}~{\rm Mpc}^{-3}$~\cite{PU90}, the inferred isotropic-equivalent UHECR luminosity is $L_{\rm UHECR} \sim {10}^{43.5}~{\rm erg}~{\rm s}^{-1}$ (regarding blazars as radio-loud AGN pointing toward us).  This is much smaller than the cosmic-ray luminosity required for explaining extreme TeV blazars, and implies that those distant radio-loud AGN with hard VHE spectra should be rarer and more powerful in cosmic rays than nearby AGN responsible for the observed UHECRs.  For 1ES 0229+200, the single-source flux is $\sim 10$~\% of the observed UHECR flux so that the anisotropy can be used as a useful probe. 
 
We demonstrated the importance of structured EGMFs that help isotropize the trajectories of UHECRs, though the EGMF strengths are still uncertain.  In addition, there are other causes that can diminish the beaming of UHECRs and resulting cascade fluxes.  
One arises from plasma instabilities induced by cosmic rays~(Murase et al. 2011, in preparation).  Second, radio lobes of powerful radio-loud AGN, like in the case of Cen A with $B \sim 1~\mu$G, would also isotropize UHECRs~\cite{der09}, as might radio bubbles from the jets of typical FR-I radio galaxies and aligned counterparts.  These magnetic fields seem relevant in order that cosmic rays from relativistic jets of radio-loud AGN to contribute to the observed flux of UHECRs.  
Indeed, for nearby radio-loud AGN, the UHECRs must be significantly isotropized, since there is no blazar (i.e., aligned radio-loud AGN) within $\sim 100$~Mpc\footnote{In other words, the ``apparent'' UHECR source density indicated from analyses of auto-correlation satisfies $n_s \gtrsim {10}^{-5}~{\rm Mpc}^{-3}$~\cite{kw08,ts09}, which is larger than the local blazar number density, $n_s \sim {10}^{-6.5}~{\rm Mpc}^{-3}$.  If UHECRs are isotropized rather than beamed, one may compare it to the local FR-I galaxy density, $n_s \sim {10}^{-4}~{\rm Mpc}^{-3}$~\citep[e.g.,][]{PU90}, which is consistent with the lower limit on the apparent UHECR source density.  Then, the inferred UHECR luminosity per source is typically $L_{{\rm UHECR},j} \sim {10}^{41}~{\rm erg}~{\rm s}^{-1}$.} and no evidence of cross-correlation with nearby blazars such as Mrk 501 and Mrk 421~\citep{der09}.  The isotropic-equivalent UHECR luminosity (at the source) $L_{\rm UHECR} \gtrsim {10}^{45}~{\rm erg}~{\rm s}^{-1}$ at $\sim 100$~Mpc will lead to overproduction of the observed UHECR spectral flux.

\section{Discussion and Summary}

In this work we studied BL Lac objects and FR-I radio galaxies as potential UHECR sources in light of recent \textit{Fermi} and imaging atmospheric Cherenkov telescope observations, and considered how future CTA, HAWC, and other high-energy $\gamma$-ray experiments might test the origin of the $\gamma$ rays from this class of blazars.  

If one accepts the standard synchrotron/SSC model for typical, FR-I galaxies and highly variable BL Lac objects that comprise the majority of VHE radio-loud AGN, the proton maximum energy is typically $\sim 1-10$~EeV unless UHE protons are produced as rare transient events, and only heavier nuclei normally reach the $\gtrsim 10^{20}$~eV energies.  In terms of the maximum energy, a heavy-ion dominated composition can be compatible with the standard SSC model because Fe nuclei can be accelerated to $\gtrsim 10^{20.5}$ eV while surviving against photodisintegration (if $\delta \gtrsim 20$; see equation 8).  
An open issue of the heavy-ion dominated composition scenario of radio-loud AGN is how the significant amount of heavy nuclei is loaded in AGN jets, which is suggested from the PAO composition results~\cite{abr10} and the observed isotropy in arrival distribution at $\sim {10}^{19}~{\rm eV}~Z_{1.5}^{-1} E_{A,20.5}$~\cite{abr11}.  

On the other hand, if hadronic models are adopted for typical, FR-I galaxies and highly variable BL Lac objects, then the observed VHE emission from these objects could be proton synchrotron radiation if protons are accelerated up to $\sim {10}^{20.5}$~eV, which requires strong magnetic fields, $B' \sim 10-100$~G that could be found in the inner jets of the radio-loud AGN.  Such hadronic models can be compatible with a proton-dominated composition.  Especially for luminous blazars with spectacular flares and low-peaked BL Lac objects with scattered radiation fields, one may expect high-energy neutrinos produced in inner jets as one of the hadronic signatures~\cite[e.g.,][]{ad01}.

In either of the synchrotron/SSC or hadronic model, we mainly considered the blazar zone in the inner jet as the emission region of $\gamma$ rays.  However, some recent studies based on simultaneous radio and $\gamma$-ray observations are questioning the standard idea that the blazar region is located near the AGN core~\citep[e.g.,][]{mar10}.  For example, in the case of 3C 345, Schinzel et al. (2012) proposed that the emitting region is located at $\sim 23$~pc along the jet. 

Extreme TeV blazars, in sources like 1ES 0229+200, 1ES 0347-121, H 2346-309, and 1ES 1101-232~\citep{nv10}, are extreme both in their deabsorbed TeV spectra and their quiescent, non-blazar-like behavior.  
Their hard source spectra can be explained by $\gamma$ rays that are produced either via electronic SSC or hadronic processes in inner jets, but hard EBL-deabsorbed VHE spectra typically require extreme source parameters or a special setup~\cite{tav11,zac11}.  
An alternative interpretation for the extreme blazars, whose variability is slow or absent, comes from intergalactic cascade emissions.  $\gamma$-ray-induced or UHECR-induced cascaded emissions can make slowly variable or almost non-variable components provided that the IGMF is weak enough, and the former may be seen as a slowly variable pair-echo component, following a more rapidly variable component.   

We examined these possibilities with numerical calculations, taking into account effects of structured EGMFs in filaments and clusters, and demonstrated that the structured EGMFs would play an important role on the proton-induced cascade emission, and that cosmic rays are significantly isotropized especially for the maximum proton energy of $\lesssim {10}^{19}$~eV.     
In this case, rather large cosmic-ray luminosities are required in order to explain the emission from extreme TeV blazars such as 1ES 0229+200 by the cascade radiation induced by UHECRs.  
Note that adopting $E_p^{\rm max} \sim {10}^{19}$~eV is motivated by values obtained from the synchrotron/SSC modeling for typical, variable radio-loud AGN rather than extreme, non-variable blazars.  One should keep in mind that the intergalactic cascade scenario itself does not tell about $E_p^{\rm max}$.  Hence, one may consider protons with $E_p^{\rm max} \sim {10}^{20}$~eV, where strong deflections in the structured regions can be avoided.  As a result, the required UHECR luminosity to power VHE emission from extreme TeV blazars becomes more reasonable, which is also comparable to values needed by the proton synchrotron model for typical, variable TeV blazars.

There are some issues in the UHECR-induced cascade scenario that need more discussion. Even if cosmic rays are completely beamed upon entering the voids of intergalactic space, the distant and luminous extreme blazars should be more powerful cosmic-ray sources than the nearby weaker FR-I galaxies found within the GZK radius that are assumed to be responsible for the highest energy UHECRs.  To avoid requiring excessively large beaming-corrected cosmic-ray luminosities to make steady TeV radiation, it is therefore better if the UHECRs from extreme TeV blazars are beamed.  On the other hand, if BL Lac objects and FR-I galaxies are the main sources of UHECRs, then the escaping UHECRs must be significantly isotropized, because most radio galaxies within the GZK radius have misdirected jets.  This isotropization could be caused by radio bubbles or the lobes of radio galaxies.  The contrary behavior may be allowed if the distant extreme blazars from which steady emission is detected are in some ways special, e.g., their host galaxy exists in a region with a very weak EGMF.

Furthermore, note that in the case of rare transient activities with short duration $t_{\rm dur}$ and longer quiescent periods $t_{\rm qui} > {\rm max}[{\Delta T}_{\rm CR},{\Delta t}_{\rm IGV}]$ between events, the intergalactic cascade scenario for extreme blazars may not work.  The timescale during which cosmic-ray-induced cascade emission lasts for long time is increased by the factor ${\rm max}[{\Delta T}_{\rm CR},{\Delta t}_{\rm IGV}]/t_{\rm dur}$.  Then, the cosmic-ray-induced $\gamma$-ray flux is consequently reduced by the cosmic-ray and pair-echo induced extension of this emission, and the required cosmic-ray luminosity has to be correspondingly increased, which could make excessive demands on UHECR power.  Such a rare transient episode could, however, produce long-lasting though faint cosmic-ray-induced $\gamma$-ray emission that lacks associated source emission, in analogous to TeV $\gamma$-ray-induced cascade radiation~\citep[e.g.,][]{2010ApJ...719L.130N}.  
On the other hand, for repeating flaring activities with the time interval $t_{\rm qui} < {\rm max}[{\Delta T}_{\rm CR},{\Delta t}_{\rm IGV}]$, the cosmic-ray-induced cascade emission can be regarded as almost persistent due to contributions from multiple flares.
 
Observational tests of the properties of the extreme TeV blazars can reveal the radiation mechanism of BL Lac objects and FR-I galaxies and provide a clue to UHECR acceleration, despite the above potential issues.  First, continuing variability searches with VERITAS, HESS, and MAGIC, and future studies with CTA are obviously important to determine if the emission must be made in the jet.  In addition, discrimination of the UHECR-induced intergalactic cascade from the $\gamma$-ray-induced cascade and attenuated source emission is possible from measurements at $\gtrsim 1-10$~TeV energies.  Detection of high-energy photons above $25$~TeV from 1ES 0229+200 or above $\sim$~TeV from more distant blazars, which may be realized by future $\gamma$-ray detectors such as CTA and HAWC, would be compelling evidence that this kind of object is a source of UHECRs.  In addition to $\gamma$-ray observations, anisotropy searches with UHECR arrival directions provide another interesting test.
 
In this work we focused on UHECR acceleration in the inner jets of BL Lac objects and FR-I galaxies.  Note that FSRQs and FR-II galaxies are rarer but more powerful, and they can also be sources of UHECRs and neutrinos.  
Other scenarios such as shock acceleration at hot spots~\citep[e.g.,][]{tak90,rb93,th11} and cocoon shocks~\citep[e.g.,][]{nma95,omy10} are viable for these types of radio-loud AGN.  But their local number density, $n_s \sim {10}^{-7.5}~{\rm Mpc}^{-3}$, appears to be too small to avoid strong anisotropy in the local universe~\cite{ts09}.  In the heavy-ion dominated composition case, not only blazars and radio galaxies, but also radio-quiet AGN~\cite{pmm09} could be sources of UHE nuclei. 

In summary, we have considered observational implications of BL Lac objects and FR-I galaxies as steady sources of UHECRs.  Within the standard synchrotron SSC model for typical, variable BL Lac objects and misaligned counterparts, acceleration of UHE protons to energies $\gtrsim 10^{20}$~eV is unlikely, so the composition of higher-energy cosmic rays should be dominated by heavy ions within the framework of this model.  The intergalactic cascade emission has to be sub-dominant for highly variable blazars and radio galaxies, while it can play a role on the spectrum of slowly variable or non-variable objects, especially extreme TeV blazars.  If the TeV spectrum of those blazars is produced by UHECR-induced cascade emission, then structured EGMFs, which can significantly isotropize protons, increase the luminosity demands on these sources.  The intergalactic cascade emissions induced by VHE/UHE photons and UHECRs from a distant source can be distinguished by future multi-TeV observations from CTA and HAWC.  In particular, detection of $\gtrsim 25$~TeV photons from relatively low-redshift sources such as 1ES 0229+200 or $\gtrsim$~TeV photons from more distant sources would favor such objects as being sources of UHECRs.   
 
\begin{acknowledgements}
K.M. acknowledges  financial support by a Grant-in-Aid from JSPS, from CCAPP and from NRL.  
The work of C.D. is supported by the Office of Naval Research and NASA Fermi Guest Investigator grants.  The work of G.M. is supported by the National Aeronautics and Space Administration through Chandra Award Number GO2-3148A and GO8-9125A, GO0-11133X issued by the Chandra X-Ray Observatory Center.  We thank John Beacom, Adrian Biland, Alexander Kusenko, Gernot Maier, and Soebur Razzaque for discussions.  We are also grateful to the anonymous referee.  
\end{acknowledgements}


\end{document}